# Reliability and fault tolerance in the European ADS project

*Jean-Luc Biarrotte*
CNRS/IN2P3, IPN Orsay, France

**Abstract**
After an introduction to the theory of reliability, this paper focuses on a description of the linear proton accelerator proposed for the European ADS demonstration project. Design issues are discussed and examples of cases of fault tolerance are given.

## 1 Introduction

The aim of an Accelerator-Driven System (ADS) is to transmute long-lived radioactive waste in a subcritical reactor. This typically requires a continuous proton beam with an energy of 600 MeV to 1 GeV, and a current of a few milliamps for demonstrators and a few tens of milliamps for large industrial systems. Such machines belong to the category of high-power proton accelerators, with an additional requirement of unprecedented reliability levels: because of the thermal stress induced in the subcritical core, the number of unwanted beam trips should not exceed a few per year, a specification that is far above usual performance and turns the issue of reliability into the main challenge and a constant consideration in all research and development activities pertaining to this type of accelerator.

This paper describes, basically, a reference solution adopted for such a machine, a superconducting linac with a combination of redundant and fault-tolerant schemes. The focus is primarily on the MYRRHA project, led by SCK•CEN in Belgium. A short presentation of the theory of reliability is also given, inspired mainly by Refs. [1, 2].

## 2 The basic concepts of reliability

Reliability deals with the analysis of failures, their causes, and their consequences. A commonly used definition of reliability is the following: 'reliability is the probability that a system will perform its intended function under a specified working condition—i.e. without failure—for a specified period of time'. This definition makes it clear that two important aspects have to be taken into account when speaking about reliability:

– a functional definition of failure is needed: in the case of an accelerator, a failure will typically be the absence of a beam on the target, or a beam on the target with the wrong parameters;

– a period of time, or 'mission time', is needed to define the reliability level of a system: unlike the availability, which measures the mean system uptime (see Section 2.3), the reliability is very time-dependent.

### 2.1 Reliability function, failure rate, mean time to failure

Mathematically, the reliability function $R(t)$ of a system can be defined as the probability that the system experiences no failures during the time interval 0 to $t$, given that it was operating at time zero. The reliability function therefore ranges between 0 and 1, by definition. In the simple but unrealistic case where a system systematically experiences a failure at time $t_{\text{fail}}$, the reliability of the system would be $R(t < t_{\text{fail}}) = 1$ and $R(t \geq t_{\text{fail}}) = 0$. In the real world, things are of course more complex, and statistical tools need to be used to correctly model how a system experiences failures.

All of the functions commonly used in reliability engineering can be derived directly from and described by a probability density function, namely the failure density distribution. The failure density distribution $f(t)$ of a system is the probability that the system experiences its first failure at time $t$, given that the system was operating at time zero.

Once this statistical distribution is defined, it is easy to derive the failure probability $F(t)$ of the system (see Fig. 1), which is the probability that the system experiences a failure between time zero and time $t$:

$$F(t) = \int_0^t f(x)\, dx. \qquad (1)$$

The reliability function $R(t)$ can then be expressed simply as

$$R(t) = 1 - F(t) = \int_t^\infty f(x)\, dx. \qquad (2)$$

Note also that, conversely,

$$f(t) = \frac{dF(t)}{dt} = -\frac{dR(t)}{dt}. \qquad (3)$$

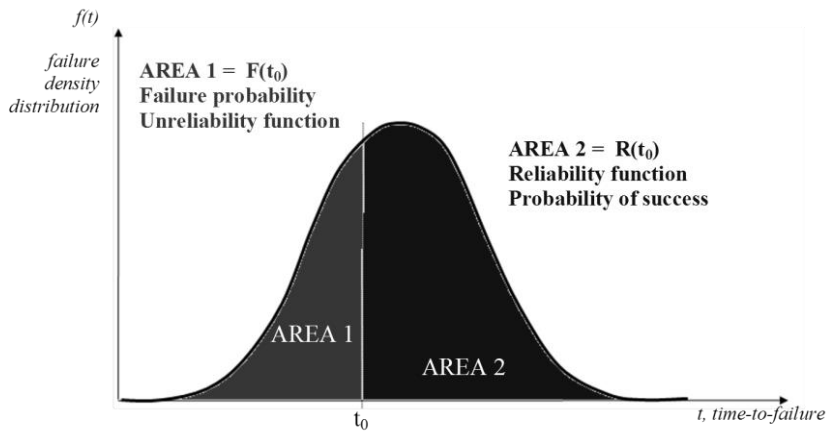

**Fig. 1:** Failure probability and reliability for a Gaussian distribution of failure density

The failure rate function $\lambda(t)$ is also an important concept; it enables the determination of the number of times the system will fail per unit time. Mathematically, it is given by

$$\lambda(t) = \frac{1}{R(t)} \frac{dF(t)}{dt} = \frac{f(t)}{R(t)}. \qquad (4)$$

This instantaneous value is also known as the hazard function. It is useful for characterizing the failure behaviour of a product, determining the allocation of maintenance crews, and planning the provision of spares. Note that Eq. (4) also leads to

$$\ln(R(t)) = -\int_0^t \lambda(t)\, dt. \qquad (5)$$

Finally, the mean time to failure (MTTF) is defined as the average time of operation of the system before a failure occurs. This value is widely used, and is often the main value of interest in characterizing the reliability of equipment. It can be computed from:

$$\text{MTTF} = \int_0^\infty t f(t)\, dt. \tag{6}$$

The MTTF is used for non-repairable systems. When dealing with repairable systems, it is more meaningful to speak about the mean time between failures (MTBF). These two metrics are identical if the failure rate of the system is constant.

## 2.2 Commonly used distributions

The reliability function, failure rate function, and mean time functions can be determined directly from the failure density distribution $f(t)$. Several distributions exist, such as the normal (Gaussian), exponential, and Weibull distributions. The simplest of these and the most commonly used—even in cases to which it does not really apply—is the exponential distribution, which can be expressed as

$$f(t) = \lambda e^{-\lambda t}, \tag{7}$$

where $\lambda$ is a constant. In this case, one can derive the following relationships:

$$F(t) = 1 - e^{-\lambda t}, \tag{8}$$

$$R(t) = e^{-\lambda t}, \tag{9}$$

$$\lambda(t) = \lambda, \tag{10}$$

$$\text{MTTF} = \frac{1}{\lambda}. \tag{11}$$

With this exponential distribution, the failure rate function $\lambda(t)$ is a constant. This means that in this case, the system does not have an ageing property. This assumption allows one to calculate the MTBF by dividing the total operating time of the system by the total number of failures encountered. Practically, this is usually valid for software systems, but most of the time, for hardware systems, the failure rate can have other, more complex shapes. A remarkable aspect of the exponential distribution is the fact that once the MTTF is known, the distribution is fully specified.

Another convenient distribution is the well-known normal distribution, given by

$$f(t) = \frac{1}{\sigma\sqrt{2\pi}} e^{-\frac{(t-\mu)^2}{2\sigma^2}}, \tag{12}$$

where $\mu$ is the mean value and $\sigma$ is the standard deviation of the distribution. In this case, the failure rate function is always increasing, and the MTTF is equal to the mean value $\mu$.

Two more powerful distributions are the lognormal and Weibull distributions. These can be applied to describe various failure processes correctly. They are given by

$$f(t) = \frac{1}{\sigma t\sqrt{2\pi}} e^{-\frac{(\ln t - \mu)^2}{2\sigma^2}}, \tag{13}$$

$$f(t) = \frac{\beta}{\eta}\left(\frac{t}{\eta}\right)^{\beta-1} e^{-\left(\frac{t}{\eta}\right)^\beta}, \tag{14}$$

respectively. For the lognormal distribution, $\mu$ is the mean value and $\sigma$ is the standard deviation; for the Weibull distribution, the two parameters are the shape parameter $\beta$ and the scale parameter $\eta$. Note that the Weibull distribution is very commonly used because it allows one to describe the ageing property of a system easily, by mixing different Weibull distributions for different stages of the life of the system, as illustrated in Fig. 2.

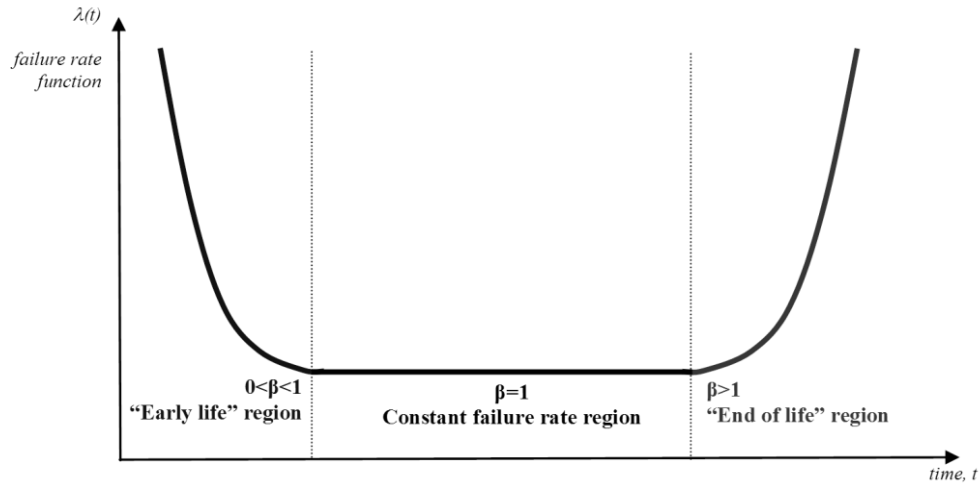

**Fig. 2:** Shape of the classical failure rate function, reconstructed using Weibull distributions, for which $\lambda(t) = (\beta/\eta)(t/\eta)^{\beta-1}$

## 2.3   Maintainability and availability

When a system fails to perform satisfactorily, repair work is normally carried out to locate and correct the fault. The system is restored to its functioning state by making an adjustment or by replacing a component, according to prescribed procedures and resources.

The maintainability is defined as the probability of isolating and repairing a fault in a system within a given period of time. Generally, exactly the same formalism as that used for reliability is used to describe maintainability. The random variable here is the time to repair, in the same way as the time to failure is the random variable in the case of reliability. One can therefore describe various aspects of maintainability using a repair density distribution and by defining a maintainability function, a repair rate function, and the mean time to repair, usually denoted by MTTR, which is the expected value of the repair time.

If one considers both the reliability (the probability that an item will not fail) and the maintainability (the probability that the item will be successfully restored after failure), then an additional metric is needed for the probability that the system is operational at a given time $t$ (i.e. either it has not failed or it has been restored after failure). This metric is called the availability and is usually denoted by $A(t)$. The availability is therefore defined as the probability that the system is operating properly when it is required for use. The availability function, which is a complicated function of time, has a simple steady-state or asymptotic expression $A$, given by

$$A = \lim_{t \to \infty} A(t) = \frac{\text{system uptime}}{\text{system uptime} + \text{system downtime}} = \frac{\text{MTBF}}{\text{MTBF} + \text{MTTR}}. \qquad (15)$$

Equation (15) is very simple and is widely used, because we are usually concerned mainly with systems running for a long time. However, one should keep in mind that until steady state is reached, the MTBF may be a function of time and the above formulation should be used cautiously.

## 2.4 Common techniques in reliability analysis

There are many techniques in reliability analysis. The most common of these are reliability block diagrams, fault tree analysis, and Monte Carlo simulations.

The reliability block diagram (RBD) is one of the conventional tools for system reliability analysis and one of the most commonly used. A major advantage of using the RBD approach is the ease of expression and evaluation of the reliability. An RBD shows the structure of the reliability of the system. It is made up of individual blocks, each block corresponding to a module or function of the system. These blocks are connected to each other by basic relationships that represent the operational configuration of the modules. The most usual connections are the following.

– *Series connection* (Fig. 3): when any module fails, the whole system fails. For a pure series system, the reliability of the system is equal to the product of the reliabilities of its constituent components.

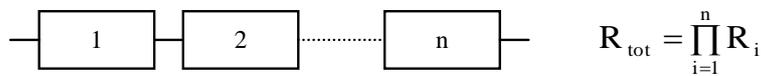

$$R_{tot} = \prod_{i=1}^{n} R_i$$

**Fig. 3:** RBD for series connection of modules with reliability $R_i$

– *Simple parallel connection* (Fig. 4): the modules are redundant, so that the system requires only one module to be operational.

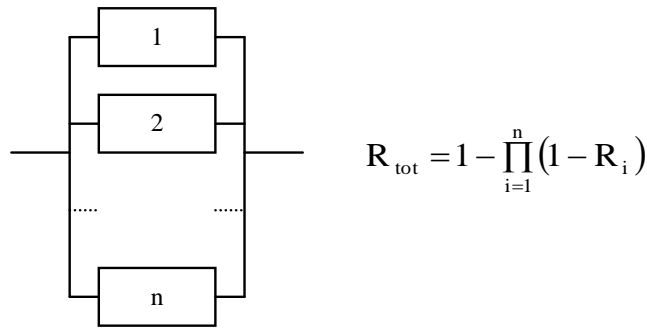

$$R_{tot} = 1 - \prod_{i=1}^{n} (1 - R_i)$$

**Fig. 4:** RBD for parallel connection of modules with reliability $R_i$

– *k-out-of-n parallel connection* (Fig. 5): the system requires at least $k$ modules out of $n$ to be operational. In this case, the redundancy is only partial.

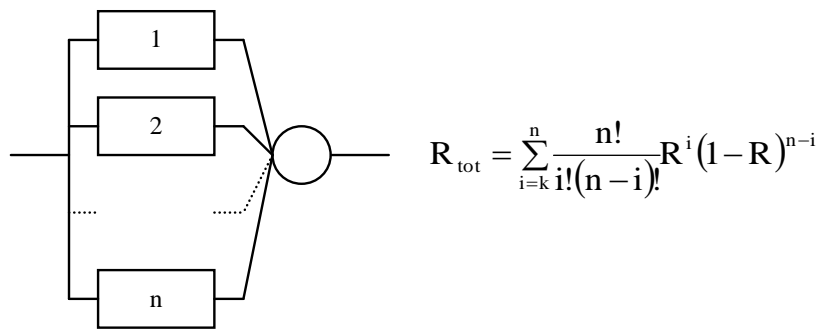

$$R_{tot} = \sum_{i=k}^{n} \frac{n!}{i!(n-i)!} R^i (1-R)^{n-i}$$

**Fig. 5:** RBD for *k*-out-of-*n* connection of modules with identical reliability $R$

Figure 6 shows a simple practical example of an RBD model for an RF system, with associated results for the reliability and availability. This example is taken from Ref. [3]. Note that this model assumes exponential failure density distributions, leading to simple derivations for the various metrics, in particular the MTBF of a series connection system:

$$\frac{1}{\mathrm{MTBF}} = \sum_i \frac{1}{\mathrm{MTBF}_i}. \tag{16}$$

Reliability and availability results for RF system (for a standard mission time of 168 h)

| Component | MTBF (1/h) | MTTR (1/h) | Failure rate (1/h) | Reliability | Availability |
|---|---|---|---|---|---|
| Transmitters | 10000 | 4 | 1.0E-4 | 0.98 | 0.99 |
| High-voltage power system | 30000 | 4 | 3.3E-5 | 0.99 | 0.99 |
| Low-level radio frequency | 100000 | 4 | 1.0E-5 | 0.99 | 0.99 |
| Power amplifiers | 50000 | 4 | 2.0E-5 | 0.99 | 0.99 |
| Power components | 100000 | 4 | 1.0E-5 | 0.99 | 0.99 |
| 1 comp./system | 5769 | 4 | | 0.97 | 0.99 |
| 60 comp./system | 96 | 4 | | 0.17 | 0.96 |

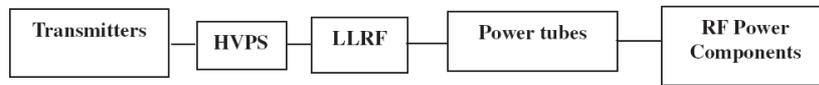

**Fig. 6:** Example of RBD analysis applied to an RF system [3] (courtesy of P. Pierini & L. Burgazzi)

Fault tree analysis, adapted from system safety analysis, is also a common tool in the application of the concepts of reliability. Whereas the reliability block diagram is mission-success-oriented, a fault tree diagram shows which combinations of component failures will result in a system failure. The fault tree diagram represents logical relationships of 'AND' and 'OR' among the various failure events, as depicted in Fig. 7. Since any logical relationship can be transformed into a combination of 'AND' and 'OR' relationships, the status of the output or top event can be derived from the status of the input events and the connections between the logical gates. A fault tree diagram can therefore describe fault propagation in a system. However, it is not always easy to describe complex systems using a fault tree formulation, especially when one wants to include aspects of repair and maintenance.

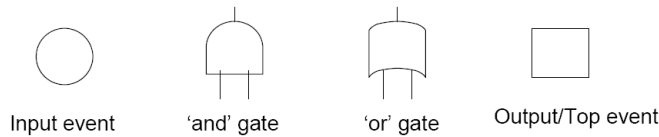

**Fig. 7:** Basic shapes used in a fault tree diagram [2]

Monte Carlo simulations are also very useful for reliability analysis and can be very powerful. In a Monte Carlo simulation, a reliability model is evaluated repeatedly using random parameter values drawn from a specific distribution. Monte Carlo simulations are often used to evaluate the MTBF for complex systems. However, they usually require the development of a customized program, and also lengthy computer runs if accurate, converging computations are desired.

## 3 The reference ADS-type accelerator

### 3.1 MYRRHA, the European ADS demonstrator project

The basic purpose of an ADS is to reduce by orders of magnitude the radiotoxicity, volume, and heat load of nuclear waste before underground storage in deep geological depositories [4]. In this context, a new research reactor, named MYRRHA (Multipurpose hYbrid Research Reactor for High-tech Applications) is being planned. It will be located at SCK●CEN, Mol, Belgium, and it is hoped that construction will start in 2015 [5]. It is designed to be able to operate in both subcritical and critical modes with the following general objectives: first, to be an experimental device to serve as a test bed for transmutation by demonstrating the ADS technology and the efficient transmutation of high-level waste; second, to be operated as a flexible, multipurpose, high-flux, fast-spectrum irradiation facility ($\Phi_{>0.75\ \text{MeV}} = 10^{15}$ n·cm$^{-2}$·s$^{-1}$); and third, to contribute to the demonstration of the Lead Fast Reactor technology, as underlined in the European Roadmap for Sustainable Nuclear Energy [6], without jeopardizing the first two objectives.

MYRRHA is composed of a proton accelerator, a spallation target, and a core with a power of ~70 MW$_{th}$ cooled by liquid lead–bismuth eutectic (LBE). To feed its subcritical core with an external neutron source, the MYRRHA facility requires a powerful proton accelerator, featuring above all a very limited number of unforeseen beam interruptions, i.e. an extremely high reliability level. The present general specifications for the proton beam are the following:

- beam energy: 600 MeV; beam energy stability: better than ±1%;
- beam pulse current: 2.5 mA, and up to 4 mA for core burn-up compensation; beam current stability: better than ±2%;
- beam time structure: continuous, with low-frequency 200 µs zero-current interruptions for on-line subcriticality monitoring of the core;
- beam footprint on the spallation target window: 'doughnut-shaped', 85 mm diameter; beam footprint stability: better than ±10%;
- beam reliability: fewer than 10 beam interruptions longer than 3 s during a three-month operation period.

Extrapolation to a 0.5 GW$_{th}$ industrial 'transmuter' leads to the following figures: 800 MeV, 20 mA proton continuous beam (total beam power 16 MW), and fewer than three beam trips per year.

### 3.2  The reliability requirement

Until now, the reliability goal for accelerators has been 'we do the best we can'. In the case of an ADS, however, reliability is a constraint for the first time. The stringent reliability requirement arises from the fact that frequently repeated beam interruptions can induce high thermal stresses and fatigue in the reactor structures, the target, and the fuel elements, with the possibility of significant damage, especially to the fuel cladding. Moreover, these beam interruptions can dramatically decrease the availability of the plant, implying plant shutdowns of tens of hours, which could quickly become unsustainable, especially for industrial transmuters.

In the case of MYRRHA, the present tentative limit for the number of allowable beam trips, 10 transients longer than 3 s per three-month operation cycle, comes from the conclusions of the EUROTRANS project [7]. This specification has been slightly relaxed compared with the initial requirements inspired by an analysis of the operation of the PHENIX reactor plant, because the MYRRHA core exhibits a fairly large thermal inertia due to the large LBE pool, and because higher margins seem to exist concerning the behaviour of the fuel and cladding during transients. This beam trip frequency nevertheless remains very significantly lower than today's reported achievements for comparable accelerators (see Fig. 8), and therefore the issue of reliability is considered as the main challenge and will be a constant consideration in all the design and R&D activities pertaining to the MYRRHA accelerator.

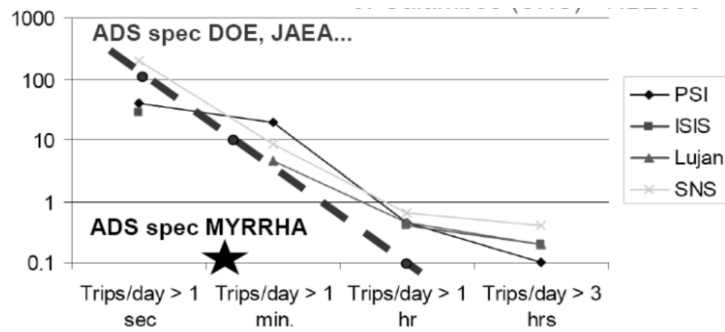

**Fig. 8:** Trip frequency vs. trip duration for high-power proton accelerators, from [8], and ADS specifications

Nevertheless, it is worth noting that several other ADS studies in Japan and the US claim beam trip limits two orders of magnitude less stringent than the MYRRHA requirements, almost compatible with the present state of the art shown in Fig. 8. In particular, a recent US Department of Energy white paper on ADS technology stated [9]: "Finding #6: Recent detailed analyses of thermal transients in the subcritical core lead to beam trip requirements that are much less stringent than previously thought; while allowed trip rates for commercial power production remain at a few long interruptions per year, relevant permissible trip rates for the transmutation mission lie in the range of many thousands of trips per year with duration greater than one second."

### 3.3 Reliability-oriented conceptual design of an ADS accelerator

The MYRRHA reliability constraint may be reformulated in the following way: the mean time between failures of the beam delivery system must be longer than ~250 h, a failure being defined as a beam trip longer than 3 s. This MTBF value is one to two orders of magnitude more demanding than what is typically achieved at present in facilities such as PSI, with a MTBF of about 1 h in 2009 [10], and the ESRF, where the operators are very much concerned about reliability and which is improving year after year, with an MTBF of more than 60 h in 2006 [11]. These figures underline the fact that reliability-oriented design practices need to be followed in the early design stage of an ADS accelerator if one wants to be able to achieve the goal.

In the accelerator context, the beam MTBF is a combination of the failure behaviour of many subsystems and sub-subsystems, all contributing fundamentally to successful beam generation. It has been shown that with such a machine configuration, an important increase in the beam MTBF may be obtained only if a single failing element does not automatically imply a global failure [3]. The key to implementing this concept of 'fault tolerance' is redundancy. Parallel redundancy can, of course, be used. It is common to use two elements for one function, as described in Section 2, but for clear economic reasons this parallel scenario has to be minimized. Serial redundancy, in contrast, replaces a missing element's functionality by retuning adjacent elements with nearly identical functionalities. This concept of serial redundancy, which is closely linked to modularity, is to be preferred when applicable.

The concept of the MYRRHA ADS machine requires a 2.4 MW proton accelerator operating in continuous mode. In principle, both cyclotrons and linear accelerators are candidates for providing such a beam. But in order to be able to implement fault tolerance and enhance the reliability figures, a modular machine, i.e. a linear accelerator ('linac'), is to be preferred. Moreover, such a solution also allows the same machine concept to be used both for the demonstrator (MYRRHA) and for long-term industrial machines, as pointed out in Ref. [12].

Thus, basically, the MYRRHA accelerator is a high-power proton accelerator with strongly enhanced reliability, but also with state-of-the-art availability (about 85%, since every beam failure will imply a rather long machine shutdown). The technical solution adopted is that of a superconducting linac, in agreement with most of the high-power accelerator projects that are in operation or to be built. The continuous operation of this type of accelerator strengthens this choice, as it ensures optimized operation costs. To implement a reliability scheme (or, equivalently, a redundancy scheme), the linac will consist of two clearly distinct sections, as illustrated in Fig. 9.

- A medium- and high-energy section (the main linac, i.e. an independently phased superconducting section), highly modular, based on individual, independently controlled accelerating cavities. In this section, serial redundancy may be applied successfully so as to yield strong fault tolerance. The function of a faulty cavity may typically be taken over by four adjacent cavities. The same concept can be applied to a faulty focusing magnet.

- A low-energy section (the injector, or linac front end), in which modularity and fault tolerance are not applicable, since the beam velocity is too low. Here the number of elements is minimized using multicell cavities, and redundancy is applied in parallel form, so that two complete injectors with fast switching capabilities are foreseen. The transition energy between the two sections has been set at 17 MeV.

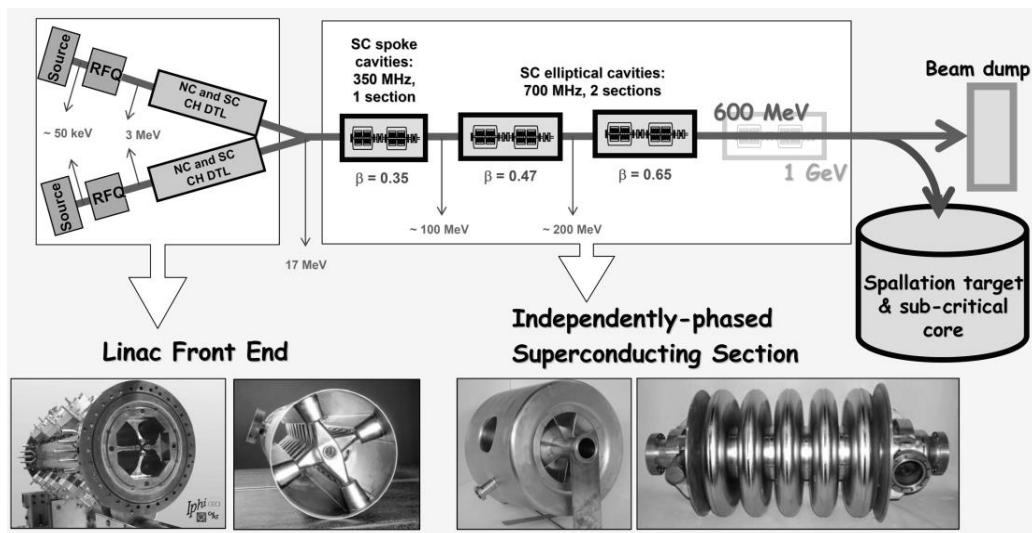

**Fig. 9:** Conceptual scheme of the MYRRHA accelerator [12] (RFQ: Radio-Frequency Quadrupole; NC: Normal Conducting; SC: SuperConducting; CH: Cross-bar H-type; DTL: Drift Tube Linac)

Throughout the design phase of the MYRRHA accelerator, which is presently ongoing with support from the FP7 European MAX project (2011–2014) [13], the following three principles therefore have been and are being followed regarding the goal of reliability:

- strong design: make it simple, avoid 'not-so-useful' complicated elements, use components far from their technological limits, and ensure 'no beam loss' operation;

- fault tolerance, and hence redundancy, with the maximum amount of serial redundancy, as already underlined, coupled with realistic fast fault recovery scenarios;

- repairability (on line where possible), coupled with a short enough MTTR and efficient maintenance schemes.

Finally, the reliability of the MYRRHA accelerator that is aimed at will only be realized if these principles are applied in every machine subsystem, including all ancillary equipment. This reliability issue will deserve continuous attention during the engineering design of all components, and during the commissioning of the components and machine.

# 4 Fault tolerance cases for MYRRHA and expected impact on reliability

## 4.1 Hot spare injector

The injector part of MYRRHA (0–17 MeV) is based on some rather unconventional solutions. These have been chosen in view of the optimal efficiency and minimized number of components that they provide, given the fact that serial redundancy cannot be applied in this section.

The present design of the injector is described in Ref. [14]; more details can be found in Ref. [13]. It is about 13 m long from the ion source exit to the entrance of the Medium Energy Beam Transport (MEBT), and is composed of four subsections:

- a 30 kV ECR (Electron Cyclotron Resonance) proton source and a 2 m long Low Energy Beam Transport (LEBT);
- a 176 MHz four-rod RFQ, 4 m long, accelerating the beam to 1.5 MeV and operating with a very conservative intervane voltage (30 kV) and Kilpatrick factor (1.0);
- two copper multicell CH-DTL structures for acceleration to 3.5 MeV;
- four superconducting multicell CH-DTL structures [15], combined in one single cryomodule, for acceleration to 17 MeV.

To increase the reliability, the philosophy here was to duplicate this 17 MeV section, providing a hot standby spare injector able to quickly resume beam operation in the case of any failure in the main injector. The fault recovery procedure is based on the use of a switching dipole magnet with a laminated steel yoke connecting the two injectors through a 'double-branch' MEBT. The injector reconfiguration process should last not more than 3 s, and is defined as follows (see Fig. 10).

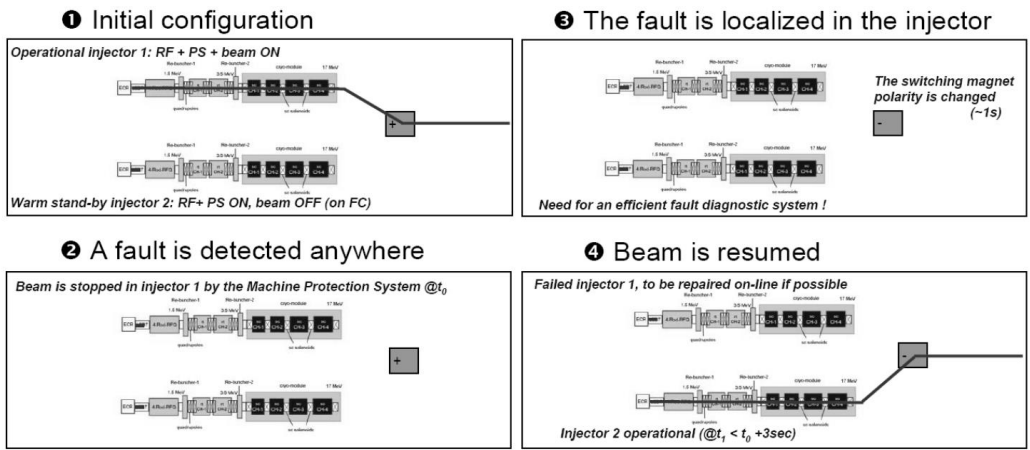

**Fig. 10:** Fault recovery scenario for the MYRRHA injector

1. In the initial configuration, one of the injectors (e.g. Injector 1) provides a beam to the main linac; the hot spare injector is also fully operational (RF on, power supplies on…), but the beam is intercepted at the source exit with a Faraday cup.

2. A fault is detected somewhere in the linac, and the beam is immediately and automatically stopped at the source exit of Injector 1 by the machine protection system.

3. The fault is localized in Injector 1 by the fault diagnostic system. The polarity of the power supply feeding the MEBT switching magnet is therefore changed.

4. Once steady state is reached, the beam is resumed using Injector 2. It is of course supposed that beam tuning of Injector 2 has been previously performed. The failed Injector 1 should then be repaired as soon as possible, during operation if possible.

## 4.2 Fault-tolerant main linac

From 17 MeV, a fully modular superconducting linac accelerates the proton beam to the final energy of 600 MeV, over a total length of about 220 m. The linac is composed of an array of independently powered 350 MHz spoke cavities and 700 MHz elliptical cavities. The design ensures a high energy acceptance and moderate energy gain per cavity, using a low number of cells and conservative accelerating gradients (with a nominal operation point of around 50 mT and 25 MV/m peak field), in order to increase the tuning flexibility as much as possible and provide a sufficient margin (about 30%) to allow the implementation of fault-recovery scenarios, as detailed below. A regular focusing-lattice scheme is used, with not-too-long cryostats and room-temperature quadrupole doublets in between. Such a scheme provides several advantages: easy maintenance and fast replacement if required, easier magnet alignment at room temperature and no fringe field issues, the possibility of providing easily accessible diagnostic ports at each lattice location, and, last but not least, nearly perfect regularity of the optical lattice (no specific beam matching is required from cryostat to cryostat). This main linac is then followed by a final transport line, composed of an achromatic triple deviation able to inject the proton beam with the specified footprint into the spallation target. More details of this design can be found in Refs. [7, 13, 16].

Because we are dealing with a non-relativistic proton beam, any fault in an RF cavity, which implies beam energy loss, will also lead to a phase slip along the linac. This will increase with distance, and thus push the beam out of the stability region: the beam will be completely lost. To recover from such RF fault conditions, the philosophy is to re-adjust the accelerating fields and phases of some of the non-faulty RF cavities to recover the nominal beam characteristics at the end of the linac, in particular the transmission, phase, and energy. A simple way to achieve this is to act on the accelerating cavities next to the failed one, as shown in Fig. 11. This 'local compensation method' has the advantage of involving only a small number of elements, and therefore of being able to compensate for multiple RF faults in different sections of the machine at the same time. A similar system based on the updating of the phase set points of all downstream superconducting cavities is being developed very successfully at the SNS [17].

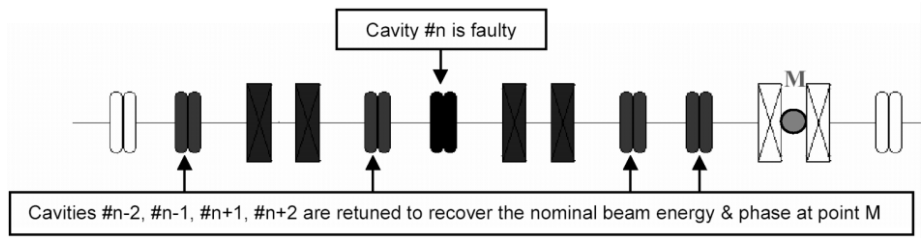

**Fig. 11:** Principle of the local compensation method

Beam dynamics simulations show that by using such a retuning method in the MYRRHA design, the nominal beam parameters at the target can always be restored with very limited additional emittance growth, given the condition that a 20–30% rise in accelerating field and RF power can be sustained in the few (four to eight) retuned elements [18]. This method is, of course, rather demanding in terms of linac length—it requires an increase of about 20%—and installed RF power budget, but is on the other hand totally in line with the ADS overdesign criterion, and is in any case required if one is to try to reach the required reliability level. This method can also be successfully applied to the failure of a focusing element. In the case of a quadrupole-based focusing scheme, it is preferable either to switch off the whole doublet in the case of a quadrupole failure—this limits the beam mismatch induced by the fault and eases the retuning of the surrounding doublets—or to use triplets.

In any case, it has to be underlined that this local compensation method can be applied successfully only if the beam velocity is high enough to avoid unsustainable defocusing effects while the beam is travelling across the drift length corresponding to the failed element. The typical limit has been found to be around $\beta = 0.15$, leading to the choice of a 17 MeV injector for MYRRHA.

Simulations of transient beam dynamics need to be performed to analyse accurately what happens to the beam during the above retuning procedures, keeping in mind that the retuning has to be performed in less than a few seconds. A new simulation tool has been developed, based on the TraceWin CEA code [19], that allows the effect of time-dependent perturbations on the beam optics to be analysed by modelling of the RF control loop. From such work [20], a reference scenario for fast on-line recovery from failures of the accelerating system has been settled on. This scenario uses the following sequence, to be performed in less than a few seconds.

1. A fault is detected somewhere in the linac, and the beam is immediately and automatically stopped at the source exit by the machine protection system.

2. The fault is localized in the RF loop of a superconducting cavity by the fault diagnostic system.

3. The field and phase set-points are updated in some RF cavities adjacent to the failed one. These set-points need to be determined previously during the commissioning phase, and possibly stored directly in digital chips in the Low Level RF (LLRF) systems.

4. To avoid any beam-loading effect, the failed cavity is detuned by a few hundred bandwidths using a suitable cold tuning system, possibly piezo-based. This sequence is the most demanding in terms of duration.

5. Once steady state is reached, the beam is resumed. The failed RF system should then be repaired as soon as possible, during operation if possible, and put back on line using a similar opposite procedure.

Details of related simulations and developments can be found in Ref. [21].

### 4.3 Expected impact on reliability

When one looks at the literature on accelerator reliability, it appears that injectors and RF systems usually represent a significant proportion of the faults that generate beam trips in operating facilities. This is the reason why the two fault cases described above have been taken into account in the early conceptual design stage of the ADS accelerator.

Based on this conceptual design, two independent integrated reliability analyses have been performed to try to estimate the number of malfunctions of the MYRRHA accelerator that could cause beam or plant shutdowns in a three-month operation cycle, and to analyse the influence of the MTBFs, the MTTRs, and the whole system architecture on the results. These studies were performed by means of a reliability block diagram analysis using the Relex© software package [3] and by means of Monte Carlo simulations using home-made software with slight differences in the hypotheses between simulations.

In both cases, the results show that a linear accelerator has high potential for reliability improvement if the system is properly designed with this objective: from about 100 unexpected beam shutdowns per three-month operation period for a classical 'all-in-series' linac, this figure falls to around three to five beam interruptions in the MYRRHA case, where a second redundant injector stage with fast switching capabilities is used, and where fault tolerance is included in the independently phased linac via fast fault recovery scenarios. Nevertheless, the absolute figures obtained remain rather questionable at present, because of the still somewhat crude modelling used for such a complex system, and because of the lack of a well-established database of component reliability figures. The development of a more accurate reliability model of the MYRRHA accelerator is therefore very much required for guidance of the engineering design. This work is presently ongoing [13], using the methodology applied in current nuclear power plants, and trying to make efficient use of existing data and models developed in the accelerator community, especially in machines rather similar to MYRRHA such as the SNS [22].

It is clear in any case that in order to reach the extremely ambitious reliability level of the MYRRHA accelerator that is desired, failure cases will have to be anticipated for all systems and subsystems (e.g. power supplies, the cryogenic system, controls, cooling systems, and vacuum systems), and suitable engineering design solutions implemented. Here again, it is extremely probable that redundancy will be a key issue, using serial redundancy as much as possible, much more often than classical duplication. Some particular fields in which very promising progress is being made in this respect are modular DC power supplies and solid-state-based RF amplifiers.

Solid-state RF amplifiers [23], for example, are totally suited to application in an ADS, with an expected MTBF of more than 50 000 h, which is clearly higher than that of classical RF tubes. These amplifiers are based on a combination of elementary modules of a few hundred watts each, providing extreme modularity and therefore inherent redundancy and flexibility towards failures. In fact, interruption of the source is not required if a failure happens in one or a few amplifier modules: operation can be still sustained with fewer modules, given that the available power remaining is sufficient. Compared with IOTs (Inductive Output Tubes) and klystrons, they also have some operational advantages such as low voltages and easy maintenance. Moreover, the continuous nature of the MYRRHA beam, and hence the absence of a peak in power demand, and the relatively low operational RF frequencies, below 1 GHz, are very compatible with such a solution.

## 5    Conclusion

In situations where accelerators are applied, one usually cares about beam availability. But with an ADS, for the first time, reliability is an additional constraint to be taken into account. Even in the case of a demonstrator, i.e. the MYRRHA machine, which requires an MTBF of about 250 h, the reliability requirement is extremely ambitious compared with the present state of the art.

Reliability models show that to be able to achieve this goal, redundancy needs to be included in all stages of the machine design in order to provide a strong level of tolerance to faults. At the level of the first conceptual design, this is achieved using a redundant injector followed by a fully modular superconducting linac with fault tolerance capabilities. At the level of the subsystems, this strategy should also be applied widely, with implementation of redundancy where possible, robust designs, and efficient maintenance strategies. Finally, once the machine is constructed, a few years of commissioning and training will be necessary to identify, repair, and optimize the weak elements so as to maximize the reliability level of the operation of the machine.

## 6    Acknowledgements

The author would like to thank all of his colleagues involved so far in these developments linked to the MYRRHA accelerator design, especially Tomas Junquera (ACS); Didier Uriot (CEA Saclay); Horst Klein, Holger Podlech, and Chuan Zhang (IAP Frankfurt); Paolo Pierini (INFN Milano); Frédéric Bouly, Christophe Joly, Alex C. Mueller, and Hervé Saugnac (IPN Orsay); and Dirk Vandeplassche (SCK•CEN). The research leading to these results has received funding from the European Atomic Energy Community (EURATOM) Fifth and Sixth Framework Programmes, and is being supported from the Seventh Framework Programme FP7/2007-2011 under grant agreement No. 269565 (MAX Project).

**References**

[1] Weibull.com, http://www.weibull.com/SystemRelWeb/blocksimtheory.htm

[2] M. Xie, K.-L. Poh, and Y.-S. Dai, *Computing System Reliability: Models and Analysis* (Springer, Berlin, 2004).

[3] P. Pierini and L. Burgazzi, *Reliab. Eng. Syst. Safety* **92** (2007) 449–463.

[4] European Technical Working Group, *The European Roadmap for Developing ADS for Nuclear Waste Incineration* (ENEA, Rome, 2001).

[5] SCK•CEN, http://myrrha.sckcen.be/

[6] SNETP, http://www.snetp.eu/

[7] J.-L. Biarrotte, A.C. Mueller, H. Klein, P. Pierini, and D. Vandeplassche, Accelerator reference design for the MYRRHA European ADS demonstrator, Proc. 25th LINAC Conf., Tsukuba, Japan, 2010.

[8] J. Galambos, T. Koseki, and M. Seidel, Commissioning strategies, operations and performance, beam loss management, activation, machine protection, Proc. 42nd ICFA Advanced Beam Dynamics Workshop, Nashville, TN, 2008.

[9] US Department of Energy, *Accelerator and Target Technology for Accelerator Driven Transmutation and Energy Production*, 2010, http://science.energy.gov/hep/news-and-resources/

[10] M. Seidel, Experience with the production of a 1.3MW proton beam in a cyclotron-based facility, Proc. 1st TC-ADS Workshop, Karlsruhe, Germany, 2010.

[11] L. Hardy *et al.*, Operation and recent developments at the ESRF, Proc. 11th EPAC Conf., Genoa, Italy, 2008.

[12] J.-L. Biarrotte *et al.*, *Nucl. Instrum. Meth. Phys. Res. A* **562** (2006) 565–661.

[13] MAX Project, http://ipnweb.in2p3.fr/MAX/

[14] C. Zhang *et al.*, From Eurotrans to MAX: new strategies and approaches for the injector development, Proc. 2nd IPAC Conf., San Sebastian, Spain, 2011.

[15] F. Dziuba *et al.*, *Phys. Rev. Spec. Top. Accel. Beams* **13** (2010) 041302.

[16] H. Saugnac *et al.*, High energy beam line design of the 600MeV 4mA proton linac for the Myrrha facility, Proc. 2nd IPAC Conf., San Sebastian, Spain, 2011.

[17] J. Galambos *et al.*, A fault recovery system for the SNS superconducting cavity linac, Proc. 23rd LINAC Conf., Knoxville, TN, 2006.

[18] J.-L. Biarrotte *et al.*, Beam dynamics studies for the fault tolerance assessment of the PDS-XADS linac design, Proc. 9th EPAC Conf., Lucerne, Switzerland, 2004.

[19] CEA, http://irfu.cea.fr/Sacm/logiciels/index3.php

[20] J.-L. Biarrotte and D. Uriot, *Phys. Rev. Spec. Top. Accel. Beams* **11** (2008) 072803.

[21] F. Bouly *et al.*, LLRF developments toward a fault tolerant Linac scheme for ADS, Proc. 25th LINAC Conf., Tsukuba, Japan, 2010.

[22] G. Dodson, The SNS reliability program, Proc. 3rd Accelerator Reliability Workshop, Cape Town, South Africa, 2011.

[23] M. Di Giacomo, Solid state RF amplifiers for accelerator applications, Proc. 23rd PAC Conf., Vancouver, Canada, 2009.